{
\documentclass[showpacs,amssymb,floatfix,pre,aps,notitlepage,preprint]{revtex4-1}

\usepackage{dcolumn}   
\usepackage{bm}        
\usepackage{amssymb}   

\usepackage{amssymb,amsfonts,bm}
\usepackage{graphics,graphicx}
\usepackage{colordvi,amsmath,epsfig,color}
\usepackage[activeacute,english]{babel}
\bibliographystyle{prsty}
\newcommand{\sig}{\boldsymbol{\hat{\sigma}}}
\providecommand{\abs}[1]{\lvert#1\rvert}

\begin{document}

\title{Homogeneous dynamics in a vibrated granular monolayer}

\author{P. Maynar}
\author{M. I. Garc\'ia de Soria}
\author{J. Javier Brey}

\affiliation{F\'{\i}sica Te\'{o}rica, Universidad de Sevilla,
Apartado de Correos 1065, E-41080, Sevilla, Spain}

\date{\today}

\begin{abstract}
A simple model of a vibrated granular monolayer is studied. It
consists of inelastic hard spheres confined between two
parallel hard plates separated a distance smaller than twice the
diameter of the particles. Both walls are elastic and one of them is
vibrating in a sawtooth way. For low densities, a kinetic equation is 
proposed from which closed evolution equations for the horizontal and
vertical temperatures are derived assuming spatial homogeneity and
that the system is very thin. An excellent agreement between the
theoretical predictions and Molecular Dynamics simulation results is
obtained both, for the stationary values and for the dynamics of the
temperatures. 
\end{abstract}

\maketitle

\section{Introduction}

A granular system is an ensemble of macroscopic particles, grains, whose
interactions are dissipative. This means that, when two particles
interact, part of the kinetic energy of the center of mass of the two
particles is transferred to another internal degree of freedom. 
Granular matter is ubiquitous
in Nature: from sand dunes to interstellar dust or planetary rings,
and they are also relevant because of its technological
applications \cite{c90}. From a theoretical point of view, granular systems are
specially interesting because, due to its dissipative character, they
are intrinsically out of equilibrium. 
A granular system can be fluidized by injecting energy using some kind 
of forcing such as vibrating walls or applying a shear.  
In this
so called fast flow regime, the dynamics is similar to that of a
normal fluid as it is, basically, a sequence of
binary collisions followed by free streaming of the grains (assuming
that the medium in which they are immersed does not affect appreciably
its movement). Due to this reminiscence to normal fluids, kinetic
equations have been used to study these 
situations and they have been proved to describe correctly the
dynamics of the system \cite{g03, at06}. In particular, hydrodynamic equations have
been derived in the free cooling case from the Boltzmann or
Enskog equations \cite{bdks98, gd99}, finding explicit expressions for the
transport coefficients. Moreover, hydrodynamic equations 
describe a variety of symmetry-breaking instabilities such as
phase-separation instability \cite{lms02, brmg02}, oscillatory instability
\cite{km04}, or thermal granular convection \cite{km03} to mention but a
few. 

A prototypical example of granular system in the fluidized regime is
an ensemble of grains inside a box in which one of the walls,
typically the one at the bottom, vibrates injecting energy into the
system. In many cases, a stationary state is reached in the long time
limit in which the energy lost in 
collisions is compensated by the energy injected by the wall. 
In the last two decades the case of a monolayer of identical spherical
grains on  a horizontal plate that is
vertically vibrated has been widely studied (see, for example, the reviews \cite{mvprkeu05,
ms16}).  
The advantage of this
kind of experimental setup with respect to the ``multilayer'' case is
that, when the height of
the system is smaller than twice the diameter of the particles, the
particles do not jump over each other and it is possible
experimentally to follow the motion of all the grains. In addition, there
also exists states in which the system can be
considered to be spatially homogeneous, while for wider systems there are
always gradients in the vertical direction. 
There are many variations of
this kind of experiment. Originally, the system is open from above, being gravity the
cause of the confinement \cite{ou98, lcg99}. The system can also be confined
by a top lid, the distance between plates being much smaller than the
horizontal dimensions in such a way that it can be considered
quasi-two-dimensional (Q2D) \cite{peu02, ou05, rcbhs11, crbhs12,
  cms12, nrtms14, cms15}. The bottom plate is usually 
smooth, although rough plates 
have also been used \cite{peu02}. Interestingly, they all share a common
phenomenology: for a wide range of the parameters, a spatially homogeneous
stationary state is reached but, for high enough densities, the system
develops cluster of particles and a final state is reached in which a
dense phase coexists with a more dilute and hotter fluid. The
instability depends also on the parameters describing the vibrating
wall. In addition, depending on the averaged density,  the coexistence
can be between a solid-like and a liquid-like phase \cite{ou98, lcg99, peu02, ou05,
  cms12, nrtms14, cms15}, or
between a liquid-like and a gas-like phase \cite{rcbhs11, crbhs12}. 

Some simple two-dimensional effective models
have been used to try to understand the above phenomenology. The
grains are modeled by inelastic hard disks and the wall by some kind
of homogeneous energy driving mechanism. In the so-called 
stochastic thermostat model, the particles are under the action of a
stochastic force with vanishing mean value and delta-correlated in
time variance \cite{wm96}. The particles suffer stochastic kicks that can inject
energy into the system. In the so called $\Delta$ model \cite{brs13},
the particles move freely between 
collisions, but the inelastic collision rule is modified by adding an
extra velocity, $\Delta$, to the relative motion pointing outwards in
the direction of the collision. Hence, the total kinetic energy of a 
pair of colliding particles can increase or decrease after a
collision. Although the $\Delta$ model seems to describe better the dynamics of
the monolayer in homogeneous situations \cite{bgmb14}, both models fail
to explain the phenomenology 
of the experiments. A homogeneous stationary state is always reached in the long time
limit in both cases, i.e. there is no presence of any instability
\cite{gmt13, gcv13, bbgm16}. Let us note that more complex
two-dimensional models have 
been studied in which the homogenous stationary state may be 
unstable. For example, if the stochastic force is multiplicative, in 
such a way that faster particles receive larger kicks, 
clusters of particles can arise \cite{clh00}. Another two-dimensional model that
presents phase separation consist in particles with an additional
variable that accounts for the kinetic energy stored in the vertical 
motion \cite{rsg18}. The parameter grows monotonically (following some
phenomenological law) until a collision takes
place and it is reset to zero. The collision rule depends
also on the parameter and, as in the $\Delta$ model, the total kinetic
energy of the pair of particles can be increased or decreased in a collision. Although 
interesting from a theoretical point of view, both models have the
disadvantage of depending on some unknown parameters that must be fitted.

When the plates are smooth, it is clear that energy is injected in the
vertical direction only, and that it is transferred to the horizontal
degrees of freedom via collisions between particles. 
In order to describe and understand from a microscopic point of view
this transference of energy, it is necessary to consider a 3
dimensional model. Very recently, the dynamics of an ensemble of 
elastic hard spheres confined between two parallel hard walls at rest separated a
distance smaller than twice the diameter of the particles has been studied
\cite{bmg16, bgm17}. For low densities, a closed equation for the
one-particle distribution function that takes into account
the effects of the confinement was formulated. The proposed
Boltzmann-like equation admits an H-theorem \cite{bmg16} and the
equilibrium distribution function derived from it agrees
with the one obtained by equilibrium statistical mechanics methods 
\cite{sl97}. Equations for the horizontal and vertical temperatures
were derived finding the specific form of the energy transfer terms 
and, also, an excellent agreement with Molecular Dynamics (MD)
simulations \cite{bgm17}. This success of kinetic theory to describe confined
elastic systems, stimulated the study of the model in the inelastic
case \cite{mgb19}, but with the bottom wall vibrating in a sawtooth
way, that always injects energy in the vertical direction. More
precisely, in Ref. 
\cite{mgb19} this inelastic model is introduced and phenomenological
equations for the 
vertical and horizontal temperatures are proposed valid for spatially
homogeneous states. The equations are
supposed to be valid only in the elastic 
limit because, for the energy transfer terms, the elastic value
deduced in \cite{bgm17} was taken. Remarkably, the pressure in the
horizontal plane in the
stationary state derived from the theory decays monotonically with the
density, implying the instability of the homogeneous
stationary state if the size of the system exceeds a critical size
\cite{mgb19}. In fact, MD simulation results show that, when the
homogeneous stationary state is unstable, a dense aggregate 
surrounded by a dilute hotter gas is formed. The situation is, then, similar to the
results of the experiments reported in Refs. \cite{rcbhs11, crbhs12}. For spatially
homogeneous situations, the predictions of the 
equations for the horizontal and vertical
temperatures agree very well with MD simulation results for mild
inelasticities, both for the stationary values and for the
dynamics. Out of this range, i.e. for stronger inelasticities, some
discrepancies arise. 

The objective of this work is to study from a microscopic point of
view the inelastic model introduced in Ref. \cite{mgb19} in the low
density regime. We will
follow the same lines stated in the elastic case. Concretely, the
first step is to extend the Boltzmann-like equation proposed in
\cite{bmg16, bgm17} to 
inelastic collisions, incorporating also the presence of the vibrating
sawtooth wall. The second objective is to derive from the kinetic
equation the equations for the horizontal and vertical
temperatures, assuming
spatial homogeneity, but without any restriction about the degree of the 
inelasticity. The idea is not only to extend the equations for the
temperatures proposed in Ref. \cite{mgb19} to any inelasticity, but also to have a complete
microscopic understanding of them. This study is also motivated by the
fact that the characterization of
these homogeneous states are essential for the derivation of
hydrodynamic equations for spatially inhomogeneous situations. The case
in which the two walls are elastic and the system cools down freely
will be studied elsewhere \cite{bgm19}. 

The paper is organized as
follows. In the next section the model is introduced and the kinetic 
equation is proposed. It is
the above mentioned extension of the kinetic 
equation introduced for elastic systems in Ref. \cite{bmg16} to inelastic
systems, incorporating also the vibrating wall. In
Sec. \ref{secEvEqs} the equations for the temperatures are obtained
from the kinetic equation assuming that the system is spatially
homogeneous and that the one-particle distribution
function is a Gaussian with two temperatures (the horizontal and
vertical temperatures). MD simulations results are presented and compared with
the theoretical predictions in Sec. \ref{secSimulations}. Sec. 
\ref{secConclusions} contains a summary of the results whose relevance
is discussed. Finally, the appendix report some details of the
calculations carried out along the paper.

\section{The model}\label{secModel}
\begin{figure}
\begin{center}
\includegraphics[angle=0,width=0.7\linewidth,clip]{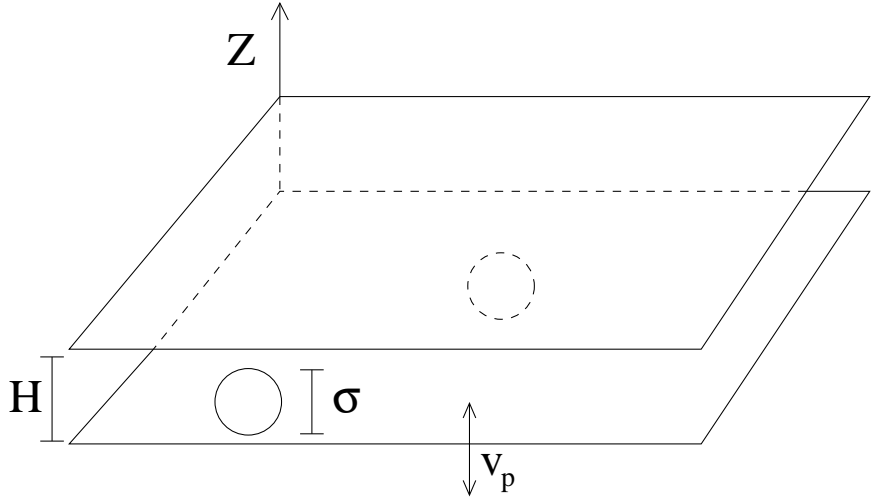}
\end{center}
\caption{Schematic representation of the model.}\label{monocapaDibujo}
\end{figure}

Let us consider an ensemble of $N$ inelastic hard spheres of mass $m$
and diameter $\sigma$ confined between two parallel rectangular shaped
plates of area $A$ separated a distance $H$. It is assumed that
$H<2\sigma$, so that particles can not jump over other particles and
the system can be considered Q2D. In the
coordinate system we will use, the plates are perpendicular to the $z$ axes and located at
$z=0$ and $z=H$, respectively. Particles move freely (gravity is not considered) 
until there is a particle-particle or particle-wall collision. When there 
is a binary encounter between two particles with velocities
$\mathbf{v}_1$ and $\mathbf{v}_2$, the postcollisional velocities,
$\mathbf{v}_1'$ and $\mathbf{v}_2'$, are  
\begin{eqnarray}
\mathbf{v}_1'\equiv b_{\sig}\mathbf{v}_1
&=&\mathbf{v}_1-\frac{1+\alpha}{2}(\sig\cdot\mathbf{v}_{12})\sig, \label{cr1}\\
\mathbf{v}_2'\equiv b_{\sig}\mathbf{v}_1
&=&\mathbf{v}_2+\frac{1+\alpha}{2}(\sig\cdot\mathbf{v}_{12})\sig. \label{cr2}
\end{eqnarray}
Here we have introduced the operator $b_{\sig}$ that replaces all
velocities $\mathbf{v}_1$ and $\mathbf{v}_2$ appearing to its right by
the postcollisional velocities,
$\mathbf{v}_{12}\equiv\mathbf{v}_1-\mathbf{v}_2$ is the relative 
velocity before the collision and $\sig$ is a unitary vector directed along the line
joining the centers of the two particles at
contact away from particle 2. The coefficient $\alpha$ is the
coefficient of normal restitution and will be considered 
to be constant (independent of the relative velocity). It goes in the
range $0\le\alpha\le 1$, being $\alpha=1$ the
elastic case. We will always consider inelastic systems,
i.e. $\alpha<1$, and periodic boundary conditions in the horizontal
directions. The top wall is elastic and at rest, so that when a particle 
collide with it simply reflects its velocity. If a particle with
velocity $\mathbf{v}$ collides with the top wall, the postcollisional
velocity is 
\begin{equation}
b_e\mathbf{v}\equiv
v_x\mathbf{e}_x+v_y\mathbf{e}_y-v_z\mathbf{e}_z, 
\end{equation}
where we have
introduced the operator $b_e$ that transforms the velocity of the
particle into its postcollisional velocity. We have also introduced
the unitary vectors in the direction 
of the axes $\{\mathbf{e}_x, \mathbf{e}_y, \mathbf{e}_z \}$. The
bottom wall is modeled by a sawtooth wall of velocity $v_p$. Within
this model, when there is a collision of a particle with the wall
(that is always at $z=0$), the
particle always sees the wall moving upwards with velocity
$v_p$. Then, if a particle with
velocity $\mathbf{v}$ collides with the bottom wall, the postcollisional
velocity is 
\begin{equation}
b_s\mathbf{v}\equiv
v_x\mathbf{e}_x+v_y\mathbf{e}_y+(2v_p-v_z)\mathbf{e}_z, 
\end{equation}
where we have introduced the corresponding operator, $b_s$. 
Note that this kind of collisions always inject energy into
the system and, as in the case of collisions with the top wall, they
conserve momentum in the direction parallel to the 
plates. Since momentum is conserved in the collisions between
particles, total horizontal momentum is a constant of the motion.
Let us also remark that, in the model, the parameter $v_p$ can always
be scaled. In effect, let us consider two ``trajectories'' of the
system, one generated by the initial conditions for the velocities,
$\{\mathbf{v}_i^{(1)}(0)\}_{i=1}^N$ and a given velocity of the
wall, $v_p$, and the other generated by $\{\mathbf{v}_i^{(2)}(0)=K
\mathbf{v}_i^{(1)}(0)\}_{i=1}^N$ and $Kv_p$ with $K$ a given
constant. As the collision rules are linear in the velocities, the
sequence of collisions is the same in both situations and $\{\mathbf{v}_i^{(2)}(t)=K
\mathbf{v}_i^{(1)}(t)\}_{i=1}^N, \forall t$. Hence, the parameter
$v_p$ will just fix the energy scale. A scheme of the model is shown
in Fig. \ref{monocapaDibujo}

The introduced model is a \emph{minimal} model to study the experimental
situations described in the previous section. Only the essential ingredients are
retained: confinement, inelasticity of collisions and energy injection
through a vibrating wall. Other aspect such as gravity, friction with
the walls, inelasticity of the walls, or friction between particles to
mention but a few, are not considered. In any case, the model conditions
are expected to hold under some well-defined physical situations,
i.e. kinetic energy of the particles much bigger than the maximum potential energy associated
to gravity, $mg(H-\sigma)$ with $g$ being the gravity acceleration, and
smooth enough particles and 
walls. Perhaps, the most crude aspect of the model is that only one of
the walls is vibrated (the sawtooth wall), while in the experiments 
the \emph{whole} box is vibrated sinusoidally. 

\begin{figure}
\begin{center}
\includegraphics[angle=0,width=0.9\linewidth,clip]{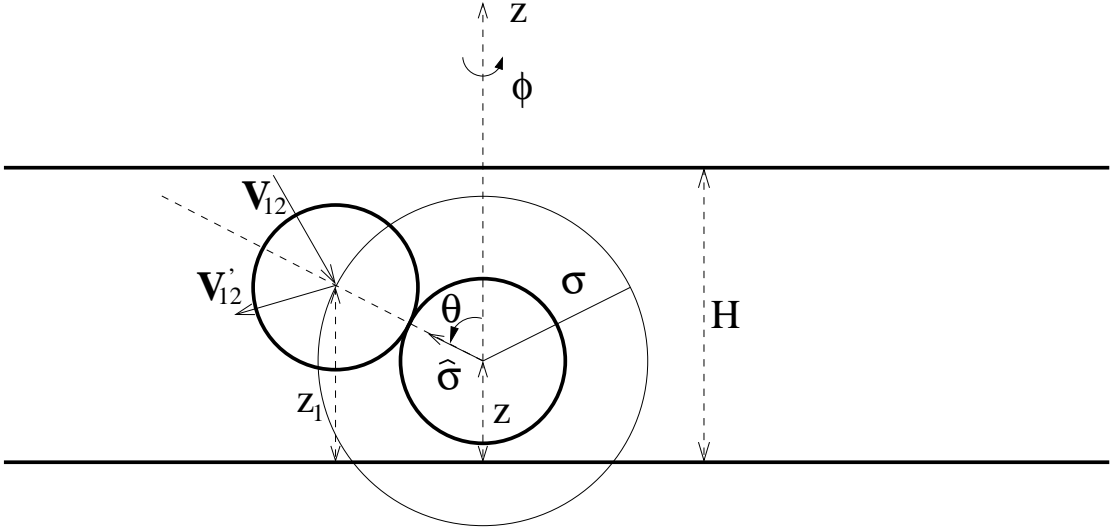}
\end{center}
\caption{Collision between two inelastic hard spheres in a Q2D
  system.}\label{monoFig}
\end{figure}

In the following, a kinetic theory description will be assumed, i.e. a
closed description in terms of the one-particle distribution function,
$f(\mathbf{r},\mathbf{v},t)$, defined as usual as the averaged density
of particles with positions around $\mathbf{r}$ and velocities around
$\mathbf{v}$ at time $t$. As said, a Boltzmann-like equation
describing the dynamics of $f$ 
of a system of  elastic hard spheres confined between two parallel elastic plates
has been proposed in Refs. \cite{bmg16, bgm17}. The generalization of the
equation to the present model is obtained by modifying the collision
rule of both the particle-particle and particle-wall collisions. 
The derivation of the equation follows standard
arguments \cite{resibois, mcLennan, dvb77}; the time evolution of the one-particle
distribution function can be decomposed in a free-streaming part, a
collisional contribution that takes into account collisions between
particles, and a wall contribution that takes into account the
collisions between the particles and the walls. In the low density limit, the collisional
term can be written in terms of $f$ by
assuming \emph{molecular chaos}, i.e. there are not velocity
correlations between the particles that are going to collide, and the equation reads
\begin{equation}\label{ecBC}
\left(\frac{\partial}{\partial
    t}+\mathbf{v}_2\cdot\frac{\partial}{\partial\mathbf{r}}\right)
f(\mathbf{r},\mathbf{v}_2,t)=J_z[f|f]+L_Wf(\mathbf{r},\mathbf{v}_2,t).  
\end{equation}
Here $J_z$ is the collisional contribution 
\begin{equation}\label{col1}
J_z[f|f]=\sigma^2\int
d\mathbf{v}_1\int_{\Omega(z)}d\mathbf{\sig}\abs{\mathbf{v}_{12}\cdot\sig}[
\Theta(\mathbf{v}_{12}\cdot\sig)\alpha^{-2}b_{\sig}^{-1}
-\Theta(-\mathbf{v}_{12}\cdot\sig)] f(\mathbf{r}+\sigma_z\mathbf{e}_z,\mathbf{v}_1,t)
f(\mathbf{r},\mathbf{v}_2,t),  
\end{equation}
where we have introduced the Heaviside step function, $\Theta$, the
operator $b_{\sig}^{-1}$ that replaces all velocities appearing to its
right by the precollisional velocities, $\mathbf{v}_1^*$ and
$\mathbf{v}_2^*$, 
\begin{eqnarray}
\mathbf{v}_1^*\equiv b_{\sig}^{-1}\mathbf{v}_1&=&\mathbf{v}_1-\frac{1+\alpha}{2\alpha}
(\sig\cdot\mathbf{v}_{12})\sig,  \\ 
\mathbf{v}_2^*\equiv b_{\sig}^{-1}\mathbf{v}_2&=&\mathbf{v}_2+\frac{1+\alpha}{2\alpha}
(\sig\cdot\mathbf{v}_{12})\sig,  
\end{eqnarray}
the $z$ component of the vector
$\boldsymbol{\sigma}\equiv\sigma\sig$, $\sigma_z$, and the region of
integration of $\sig$, $\Omega(z)$, that depends on the
confinement. In spherical coordinates, $d\sig=\sigma^2\sin\theta
d\theta d\phi$, where $\theta$ and $\phi$ are the polar and azimuthal angles
respectively (see Fig. \ref{monoFig}) and the set $\Omega$ can be parametrized as 
\begin{equation} \label{omegaPar}
\Omega(z)=\left\{(\theta, \phi)|\theta\in\left(\frac{\pi}{2}-b_2(z),
  \frac{\pi}{2}+b_1(z)\right), \phi\in(0,2\pi)\right\}, 
\end{equation}
with
\begin{eqnarray}
b_1(z)&=&\arcsin\left(\frac{z-\sigma/2}{\sigma}\right)\\
b_2(z)&=&\arcsin\left(\frac{H-z-\sigma/2}{\sigma}\right). 
\end{eqnarray}
Finally, the wall contributions is \cite{dvb77}
\begin{equation}
L_Wf(\mathbf{r},\mathbf{v},t)=[\delta(z-H+\sigma/2)L_e+\delta(z-\sigma/2)L_s]
f(\mathbf{r},\mathbf{v},t)
\end{equation}
with
\begin{eqnarray}
L_sf(\mathbf{r},\mathbf{v},t)&=&[\Theta(v_z-2v_p)\abs{2v_p-v_z}b_s
-\Theta(-v_z)\abs{v_z}]f(\mathbf{r},\mathbf{v},t), \\
L_ef(\mathbf{r},\mathbf{v},t)&=&[\Theta(-v_z)\abs{v_z}b_e
-\Theta(v_z)v_z]f(\mathbf{r},\mathbf{v},t). 
\end{eqnarray}
Let us
also mention that Eq. (\ref{ecBC}) can be directly derived from the
first equation of the BBGKY hierarchy by doing the following
approximation for the two-particle distribution function, 
$f_2(\mathbf{r}_1,\mathbf{v}_1,\mathbf{r}_2,\mathbf{v}_2,t)$, 
\begin{equation}\label{molCh}
f_2(\mathbf{r}+\boldsymbol{\sigma},\mathbf{v}_1,\mathbf{r},\mathbf{v}_2,t)\approx
f(\mathbf{r}+\sigma_z\mathbf{e}_z,\mathbf{v}_1,t)f(\mathbf{r},\mathbf{v}_1,t), 
\end{equation}
for precollisional velocities, i.e. $\mathbf{v}_{12}\cdot\sig<0$, as
was done in \cite{mgb18} for elastic hard spheres. 

Let us examine the main differences of Eq. (\ref{ecBC}) with the ``traditional'' Boltzmann
equation (without confinement). In the latter, 
the integration in $\sig$ is over all the solid angles because, at any
position, collisions with any orientation are possible. In
contrast, in Eq. (\ref{ecBC}) this is not the case because, due to the
confinement, given a
tagged particle, only collisions with the orientation
$\sig\in\Omega(z)$ are possible. Otherwise, particle $1$ would not
fulfill the constrain to be between the two walls. In addition, in the traditional
Boltzmann equation it is assumed that the one-particle distribution
function does not vary appreciably over distances of the order of $\sigma$, so that 
$f(\mathbf{r}+\boldsymbol{\sigma},\mathbf{v},t)\approx
f(\mathbf{r},\mathbf{v},t)$. On the other hand, in Eq. (\ref{ecBC}) this
approximation can not be done in the $z$ direction because an
inconsistent equation would be obtained. In effect, as $\Omega$
depends on $z$, the collisional term, $J_z$, depends also on $z$ and, due to
the kinetic equation, $f$ does depend explicitly on it. The same
occurs with the wall terms. Let us also
remark that the same two differences between the confined and
traditional Boltzmann equation are present in the elastic case. In fact, the
density profile in equilibrium, $n_e(z)$, calculated with the confined kinetic equation agrees 
very well with Molecular Dynamics simulation results \cite{bmg16}, 
consistently with the $z$ dependence of the distribution function.

%

Sometimes it is convenient to change variables in the collisional term
from the azimuthal angle, $\theta$, to the $z$ coordinate of the
particle that is going to collide with the tagged particle (see
Fig. \ref{monoFig}), $z_1$, 
\begin{equation}\label{thetaz1}
z_1=z+\sigma \cos\theta. 
\end{equation}
In these variables it is
\begin{eqnarray}\label{col2}
&&J_z[f|f]=\sigma\int
d\mathbf{v}_1\int_{\sigma/2}^{H-\sigma/2}dz_1\abs{\mathbf{v}_{12}\cdot\sig(z,z_1,\phi)}
[\Theta(\mathbf{v}_{12}\cdot\sig(z,z_1,\phi))\alpha^{-2}b_{\sig}^{-1}
-\Theta(-\mathbf{v}_{12}\cdot\sig(z,z_1,\phi))] \nonumber \\
&&f(x,y,z_1,\mathbf{v}_1,t)f(\mathbf{r},\mathbf{v}_2,t),  
\end{eqnarray}
where
\begin{equation}\label{sigmatilde}
\sig(z,z_1,\phi)=\sqrt{1-\left(\frac{z_1-z}{\sigma}\right)^2}\cos\phi\ \mathbf{e}_x+
\sqrt{1-\left(\frac{z_1-z}{\sigma}\right)^2}\sin\phi\ \mathbf{e}_y
+\frac{z_1-z}{\sigma}\ \mathbf{e}_z. 
\end{equation}
The advantage of working with $z_1$ is that the limits of integration
in the collisional term do
not depend on $z$ but, on the other hand, the dependance is translated
to $\sig$ through Eq. (\ref{sigmatilde}). 

As said, Eq. (\ref{ecBC}) implies a $z$ dependence in the one particle
distribution function. Nevertheless, for very dilute systems, it is a
good approximation to neglect this dependence, i.e. 
\begin{equation}
f(\mathbf{r},\mathbf{v},t)\approx
f(\mathbf{r}_{||},\mathbf{v},t)\equiv
\frac{1}{H-\sigma}\int_{\sigma/2}^{H-\sigma/2}dz
f(\mathbf{r},\mathbf{v},t), 
\end{equation}
where we have introduced the parallel component of a vector through 
$\mathbf{a}_{||}\equiv a_x\mathbf{e}_x+a_y\mathbf{e}_y$. In this
situation, by integrating over $z$ in Eq. (\ref{ecBC}) and replacing 
$f(\mathbf{r},\mathbf{v},t)$ by $f(\mathbf{r}_{||},\mathbf{v},t)$ in
the collisional operator, $J_z$, it is obtained
\begin{equation}\label{ecBCint}
\left(\frac{\partial}{\partial
    t}+\mathbf{v}_{2||}\cdot\frac{\partial}{\partial\mathbf{r}_{||}}\right)
f(\mathbf{r}_{||},\mathbf{v}_2,t)=\frac{1}{H-\sigma}\int_{\sigma/2}^{H-\sigma/2}dzJ_z[f|f]+(L_e+L_s)
f(\mathbf{r}_{||},\mathbf{v}_2,t), 
\end{equation}
that is a closed evolution equation for
$f(\mathbf{r}_{||},\mathbf{v},t)$. 
Of course, Eq. (\ref{ecBCint}) is fully consistent as no term depends
on $z$. The equation is simpler than Eq. (\ref{ecBC}) as there have been a
reduction in the state variables from $(\mathbf{r},\mathbf{v})$ to
$(\mathbf{r}_{||},\mathbf{v})$, i.e. the variable $z$ has disappeared
although $v_z$ remains. In the following, we will assume that the
dynamics of the system is given by Eq. (\ref{ecBCint}). 

\section{Evolution equations for the horizontal and vertical
  temperatures}\label{secEvEqs}

Let us consider spatially homogeneous states,
i.e. $f(\mathbf{r}_{||},\mathbf{v},t)=f(\mathbf{v},t)$. The objective in this
section is to derive evolution equations for the horizontal and
vertical granular temperatures, $T$ and $T_z$, that are defined as 
\begin{eqnarray} 
nT(t)&=&\frac{m}{2}\int d\mathbf{v}(v_x^2+v_y^2)f(\mathbf{v},t), \\
\frac{n}{2}T_z(t)&=&\frac{m}{2}\int d\mathbf{v}v_z^2f(\mathbf{v},t), 
\end{eqnarray}
where $n$ is the number density, $n\equiv\frac{N}{A(H-\sigma)}$, and we
have assumed that there is no macroscopic velocity field, i.e. 
$\int d\mathbf{v}\mathbf{v}f(\mathbf{v},t)=\mathbf{0}$. To
proceed, we take velocity moments in the kinetic equation. Multiplying
Eq. (\ref{ecBCint}) by $\frac{m}{2}(v_x^2+v_y^2)$ and integrating in
the velocity, it is obtained
\begin{equation}\label{m1}
n\frac{dT}{dt}=\frac{1}{H-\sigma}\int_{\sigma/2}^{H-\sigma/2}dz\int
d\mathbf{v}\frac{m}{2}(v_x^2+v_y^2)
J_z[f|f]. 
\end{equation}
Note that the walls contribution trivially vanishes since there is not
energy injection in the horizontal direction. Analogously, multiplying
Eq. (\ref{ecBCint}) by $\frac{m}{2}v_z^2$ and integrating in
the velocity, it is obtained
\begin{equation}\label{m2}
\frac{n}{2}\frac{dT_z}{dt}=\frac{1}{H-\sigma}\int_{\sigma/2}^{H-\sigma/2}dz\int
d\mathbf{v}\frac{m}{2}v_z^2J_z[f|f]+\frac{m}{2(H-\sigma)}\int
d\mathbf{v}v_z^2L_sf(\mathbf{v},t).  
\end{equation}
In this case, the top wall does not contribute (it is at rest), while the bottom wall
contribution is expressed in terms of the $L_s$ operator. 

To close equations (\ref{m1}) and (\ref{m2}), we have to 
express the velocity moments of the collisional term in terms of the
horizontal and vertical temperatures. In order to do it, 
we will assume that the distribution function is, for all
times, very closed to a Maxwellian distribution characterized by the 
temperatures in the vertical and horizontal directions, i.e. 
\begin{equation}\label{fapprox}
f(\mathbf{v},t)=\frac{n}{\pi^{3/2}w^2(t)w_z(t)}e^{-\frac{v_x^2}{w^2(t)}-\frac{v_y^2}{w^2(t)}-\frac{v_z^2}{w_z^2(t)}}, 
\end{equation}
where we have introduced the thermal velocities in the horizontal and
vertical direction, $w$ and $w_z$, through 
\begin{eqnarray}
\frac{m}{2}w^2(t)&\equiv&T(t), \\
\frac{m}{2}w_z^2(t)&\equiv&T_z(t). 
\end{eqnarray}
The validity of the approximation will be confirmed by MD simulation
results. The calculation is done in Appendix \ref{cf}, obtaining 
\begin{eqnarray}
\frac{1}{H-\sigma}\int_{\sigma/2}^{H-\sigma/2}dz\int
d\mathbf{v}\frac{m}{2}(v_x^2+v_y^2)J_z[f|f]=
\frac{2\sqrt{2\pi}(1+\alpha)n^2\sigma^2}{\epsilon}\int_0^\epsilon
  dy(\epsilon-y)(1-y^2) \nonumber\\ \label{ecTemxy}
\left\{\frac{1+\alpha}{2}\left[w^2(1-y^2)+w_z^2y^2\right]^{3/2}
-w^2 \left[w^2(1-y^2)+w_z^2y^2\right]^{1/2} 
\right\}, 
\end{eqnarray}
\begin{eqnarray}
\frac{1}{H-\sigma}\int_{\sigma/2}^{H-\sigma/2}dz\int
d\mathbf{v}\frac{m}{2}v_z^2J_z[f|f]=
\frac{2\sqrt{2\pi}(1+\alpha)n^2\sigma^2}{\epsilon}\int_0^\epsilon
  dy(\epsilon-y)y^2 \nonumber\\ \label{ecTemz}
\left\{\frac{1+\alpha}{2}\left[w^2(1-y^2)+w_z^2y^2\right]^{3/2}
-w_z^2 \left[w^2(1-y^2)+w_z^2y^2\right]^{1/2}
\right\}, 
\end{eqnarray}
where we have introduced the dimensionless parameter 
$\epsilon\equiv\frac{H-\sigma}{\sigma}$. Although the above integrals
can be evaluated exactly, their expressions are very long and we
prefer to leave them in the more compact given form. The wall
contribution is also evaluated in the Appendix, obtaining
\begin{equation}
\frac{m}{2(H-\sigma)}\int
d\mathbf{v}v_z^2L_sf(\mathbf{v},t)=\frac{nv_pT_z}{\epsilon\sigma}, 
\end{equation}
that coincides with the exact result derived in \cite{br09} (it is
$v_p$ times the pressure of the granular system just  above the
vibrating wall in the direction perpendicular to it). 

In the following, we will perform an expansion of the collisional
terms to third order in $\epsilon$. The reason is that, in this case,
the obtained expressions are easier to handle and the several
terms can be understood intuitively in a simple way. To this order, the equations are
\begin{eqnarray}
\frac{dT}{dt}&=&\sqrt{\pi}(1+\alpha)\epsilon
  n\sigma^2\sqrt{\frac{T}{m}}\left[-(1-\alpha)T
+\epsilon^2\left(-\frac{5\alpha-1}{12}T+\frac{3\alpha+1}{12}T_z\right)\right], \label{ecT}
  \\
\frac{dT_z}{dt}&=&\frac{2}{3}\sqrt{\pi}(1+\alpha)\epsilon^3
  n\sigma^2\sqrt{\frac{T}{m}}\left(\frac{1+\alpha}{2}T-T_z\right)
+\frac{2v_pT_z}{\epsilon
 \sigma}. \label{ecTz}
\end{eqnarray}
Let us briefly analyze the structure of the equations. First, let us
mention that, as the equations are obtained as an expansion in powers
of $\epsilon$, they are only valid for very thin systems. Moreover,
the considered third order in $\epsilon$, is the lowest order
consistent with the existence of a non trivial stationary state. In
effect, neglecting the 
$\epsilon^3$ terms, the only stationary state is the one with
vanishing temperatures. The first order in $\epsilon$ term in
Eq. (\ref{ecT}) is, essentially, the cooling term due to the
inelasticity of the collisions. Actually, it coincides with the
cooling term of a free evolving hard disks system in the Gaussian
approximation \cite{vne98}. The $\epsilon^3$ terms in the equations
describe energy transfer from the vertical to the horizontal degrees of
freedom due to collisions between particles. Finally, the last term in
Eq. (\ref{ecTz}) is the energy 
injection term due to collisions of particles with the bottom wall. Hence, the
dynamics can be summarized as follows: particle-bottom wall
collisions inject energy in the vertical direction and
particle-particle collisions transfer energy form the vertical to the
horizontal directions and also dissipate it. Let us also mention that,
for small inelasticities, Eqs. (\ref{ecT}) and (\ref{ecTz}) reduce to
the ones used in Ref. \cite{mgb19}, and in the elastic case
($\alpha=1$ with $v_p=0$) to the ones of Ref. \cite{bgm17}. 

Eqs. (\ref{ecT}) and (\ref{ecTz}) present a simpler form when
expressed in the dimensionless time scale, $s(t)$, defined through  
\begin{equation}
s(t)=\sqrt{\frac{\pi}{2}}(1+\alpha)n\sigma^2\epsilon\int_0^t dt'w(t'), 
\end{equation}
that is proportional to the number of collision per particle in
the time interval $(0,t)$. In effect, let us introduce the dimensionless
temperatures 
\begin{eqnarray}
\widetilde{T}&\equiv&\frac{T}{mv_p^2}, \\
\widetilde{T_z}&\equiv&\frac{T_z}{mv_p^2}. 
\end{eqnarray}
The evolution equations are
\begin{eqnarray}
\frac{d\widetilde{T}}{ds}&=&-(1-\alpha)\widetilde{T}
+\epsilon^2\left(-\frac{5\alpha-1}{12}\widetilde{T}
+\frac{3\alpha+1}{12}\widetilde{T_z}\right), \label{ecT2}
  \\
\frac{d\widetilde{T_z}}{ds}&=&\frac{2}{3}\epsilon^2\left(\frac{1+\alpha}{2}\widetilde{T}
-\widetilde{T_z}\right)+\frac{2}{\sqrt{\pi}(1+\alpha)n\sigma^3\epsilon^2}
\frac{\widetilde{T_z}}{\sqrt{\widetilde{T}}},   \label{ecTz2}
\end{eqnarray}
that do not depend on $v_p$ as a consequence of the property
mentioned in Sec. \ref{secModel} that $v_p$ only sets the energy
scale. 

From Eqs. (\ref{ecT}) and (\ref{ecTz}) (or  Eqs. (\ref{ecT2}) and
(\ref{ecTz2})) the stationary temperatures, 
$T_s$ and $T_{z,s}$, can
be easily calculated. From the horizontal temperature equation, it
follows that the ratio of stationary temperatures is 
\begin{equation}\label{ecCociente}
\gamma\equiv\frac{T_{z,
    s}}{T_s}=\frac{12(1-\alpha)+(5\alpha-1)\epsilon^2}{(3\alpha+1)\epsilon^2}, 
\end{equation}
that is density independent. The stationary horizontal temperature is
smaller than the stationary vertical temperature for $0\le\alpha<1$
and $0<\epsilon<1$, with equipartition holding in the 
elastic limit, i.e. $\lim_{\alpha\to 1}\gamma=1$. 
From the vertical temperature equation, the stationary horizontal
temperature is obtained as 
\begin{equation}\label{ecTempSt}
T_s=\left[\frac{3\gamma}{\sqrt{\pi}(1+\alpha)\left(\gamma-\frac{1+\alpha}{2}\right)
\epsilon^3\tilde{n}\sigma^2}\right]^2mv_p^2, 
\end{equation}
where the effective two-dimensional density,
$\tilde{n}\equiv\frac{N}{A}$, has been introduced. As the dimensionless parameters
$\epsilon$ and $\tilde{n}\sigma^2$ are supposed to be small, the
thermal horizontal and vertical velocities are much bigger than the
velocity of the wall, $v_p$. This can be intuitively understood since
the thinner and the more dilute is the system, the bigger is the ratio
between the particle-wall collisions and particle-particle
collisions. As the collisions with the sawtooth wall always inject
energy, the temperature increases when $\epsilon$ and/or
$\tilde{n}\sigma^2$ decrease. 

\begin{figure}
\begin{center}
\includegraphics[angle=0,width=0.7\linewidth,clip]{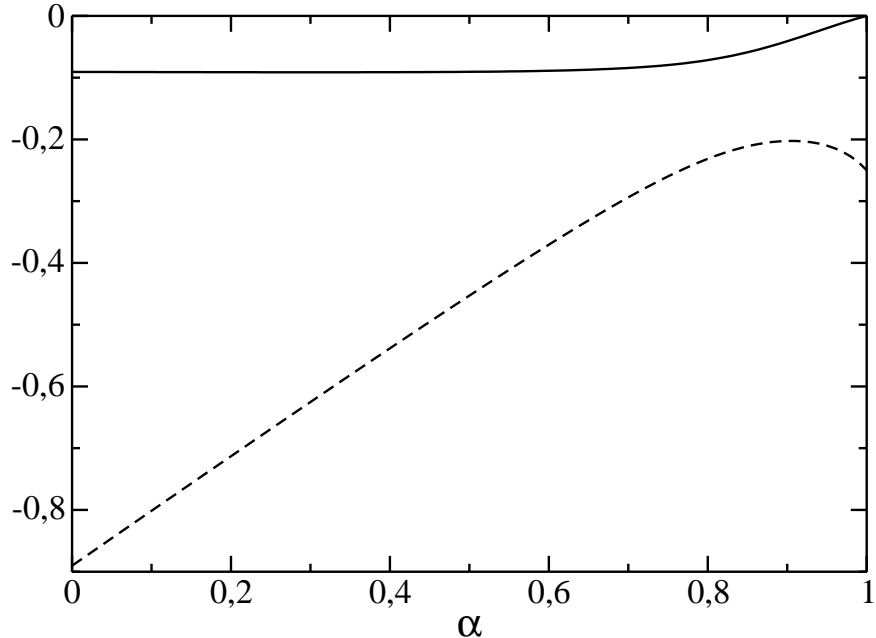}
\end{center}
\caption{Eigenvalues of the matrix $M$, $\lambda_1$ (solid
line) and $\lambda_2$ (dashed line), for $\epsilon=0.5$ as a function of the
inelasticity.}\label{autovFig}
\end{figure}

Let us consider now situations in which both temperatures are close to
their respective stationary values. Then, to linear order, the
deviations $\delta T\equiv T-T_s$ and $\delta T_z\equiv T_z-T_{z,s}$
obey the following set of linear differential equations
\begin{equation}\label{linearizedEqs}
\frac{d}{ds}\left[\begin{array}{c}
\delta T\\
\delta T_z
\end{array}\right]=M \left[\begin{array}{c}
\delta T\\
\delta T_z
\end{array}\right], 
\end{equation}
where we have introduced the matrix
\begin{equation}
M=\left[\begin{array}{cc}
-1+\alpha-\frac{5\alpha-1}{12}\epsilon^2 & \frac{3\alpha+1}{12}\epsilon^2\\
\left(\frac{1+\alpha}{2}-\frac{\gamma}{3}\right)\epsilon^2 &
                                                             -\frac{1+\alpha}{3\gamma}\epsilon^2 
\end{array}\right]. 
\end{equation}
The solution of the system is 
\begin{equation}
\left[\begin{array}{c}
\delta T(s)\\
\delta T_z(s)
\end{array}\right]=\sum_{i=1}^2\mathbf{v}_1\cdot \left[\begin{array}{c}
\delta T(0)\\
\delta T_z(0)
\end{array}\right]\mathbf{u}_i e^{\lambda_is},  
\end{equation}
where $\{\lambda_1, \lambda_2\}$, $\{\mathbf{u}_1,\mathbf{u}_2\}$ and 
$\{\mathbf{v}_1,\mathbf{v}_2\}$ are the eigenvalues, right
eigenfunctions and left eigenfunctions of $M$ respectively. The
eigenvalues are always negative so that the system of differential
equations given by Eq. (\ref{linearizedEqs}) is linearly
stable. In Fig. \ref{autovFig} the eigenvalues, $\lambda_1$ (solid
line) and $\lambda_2$ (dashed line), are plotted for $\epsilon=0.5$ as a function of the
inelasticity. It is found that the two eigenvalues are always separated, one of them,
$\lambda_1$, being the slowest and vanishing in the elastic limit. The
same kind of behavior is obtained for a wide range of the values of
the parameters although, for $\epsilon\sim 0.8$ the eigenvalues cross each other at
$\alpha\sim 0.4$. In any case, it is
not clear that for such a height Eqs. (\ref{ecT}) and (\ref{ecTz})
describe correctly the dynamics of the system. 
The fact that the eigenvalues are always well separated implies that
there is a time scale, that will be called ``homogeneous
hydrodynamic'' time scale, in which the dynamics is governed by the
slowest mode, $\lambda_1$. In this regime the two temperatures are
related through 
\begin{equation}
\delta T_z\approx q\delta T
\end{equation}
with
\begin{equation}\label{ecP}
q\equiv\frac{u_{11}}{u_{12}}=\frac{12(\lambda_1+1-\alpha)+(5\alpha-1)\epsilon^2}
{(3\alpha+1)\epsilon^2} . 
\end{equation}
In the next section, we
will see that the homogeneous hydrodynamic regime is not an exclusive
characteristic of the linear case, but that it also arises in general.

\section{Simulation results}\label{secSimulations}

\begin{figure}
\begin{center}
\includegraphics[angle=0,width=0.7\linewidth,clip]{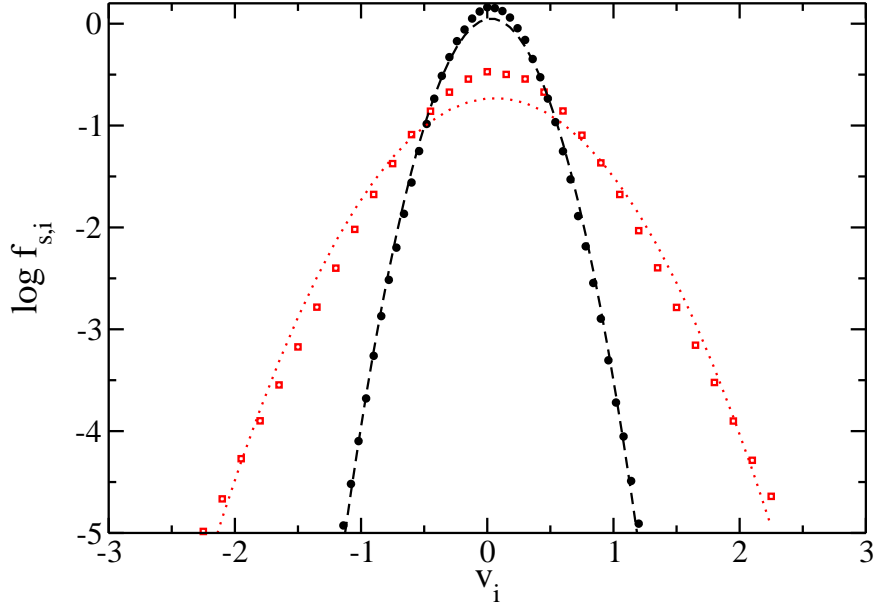}
\end{center}
\caption{Simulation results for the logarithm of
the stationary marginals velocity 
distribution functions (color online), $f_{s,x}(v_x)$ (black circles) and $f_{s,z}(v_z)$ 
(red squares) for $\epsilon=0.5$ and
$\alpha=0.8$ as a function of $v_x$ and $v_z$ respectively. The black
dashed line and point red line are the corresponding quadratic
interpolations. }\label{FigLnfdist}
\end{figure}

To check the validity of the results of the previous section, we have
carried out MD simulations of our model using
the event driven algorithm \cite{allen}. The instantaneous positions and velocities
of all the grains have been measured for different values of
the parameters, starting with 
different initial conditions. All the
simulations are run with $N=500$ and $\tilde{n}\sigma^2=0.03$, taking
the mass of the grains as the unit of mass. In most of the simulations
$v_p=0.001\left[\frac{T_0}{m}\right]^{1/2}$ with $T_0\equiv T(0)$ the
initial horizontal temperature that is taken to be unity. If not, it
is explicitly indicated. The initial condition 
was taken to be an anisotropic 
Gaussian of the form given by Eq. (\ref{fapprox}). We have checked that the
system stays for all times spatially homogeneous in the horizontal
direction and that, after a transient, 
a steady state is reached. Let us first
focus on the properties of this stationary state. In this case, all the
results are generated with one trajectory, averaging over a 
time period of about $50000$ total collisions (particle-particle and
particle-wall) per particle once the steady state is reached. In
Fig. \ref{FigLnfdist} (color online), the simulation results for the logarithm of
the stationary marginals velocity 
distribution functions, $f_{s,x}(v_x)\equiv\int
dv_y\int dv_zf_s(\mathbf{v})$ (black circles) and $f_{s,z}(v_z)\equiv\int
dv_x\int dv_yf_s(\mathbf{v})$ (red squares), are plotted for $\epsilon=0.5$ and
$\alpha=0.8$ as a function of $v_x$ and $v_z$ respectively. The black
dashed line and point red line are the corresponding quadratic
interpolations. It is found that they can
be very well fitted by Gaussians as was said in the previous
section. Similar results are found for other values of the
parameters. 

\begin{figure}
\begin{center}
\includegraphics[angle=0,width=0.7\linewidth,clip]{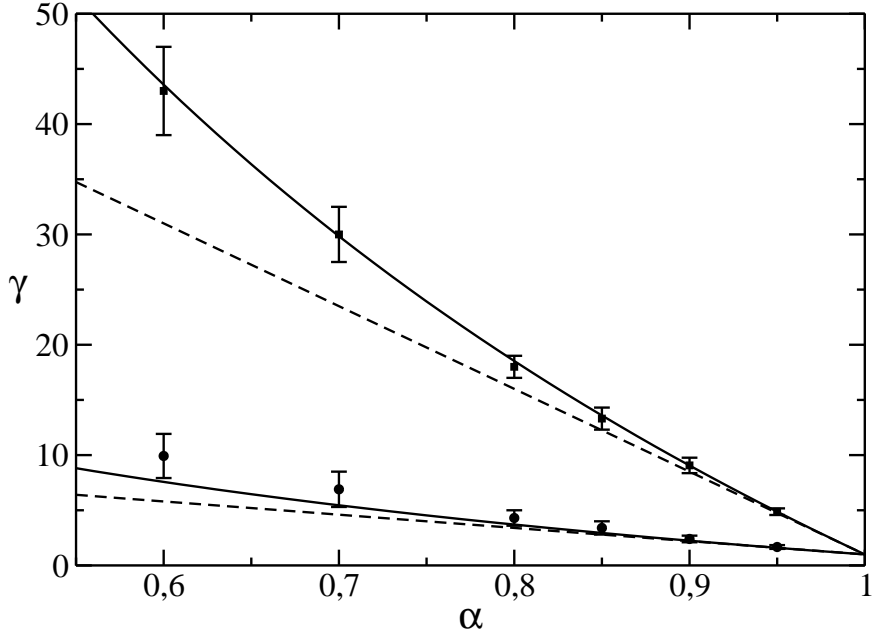}
\end{center}
\caption{ Ratio between the stationary temperatures, $\gamma$, for
$\epsilon=0.5$ (circles) and $\epsilon=0.2$ (squares) as a function of
the inelasticity. The solid lines are
the corresponding theoretical prediction given by
Eq. (\ref{ecCociente}) and the dashed lines the quasielastic
theoretical prediction of Ref. \cite{mgb19}.}\label{FigCociente}
\end{figure}

In Fig. \ref{FigCociente} the ratio between the stationary
temperatures, $\gamma\equiv\frac{T_{z,s}}{T_s}$, is plotted for
$\epsilon=0.5$ (circles) and $\epsilon=0.2$ (squares), and the coefficient of normal
restitution in the range $0.6\le\alpha\le 0.95$. The error bars have
been calculated from the dispersion of the temperatures measured once
the stationary state has been reached. The solid lines are
the corresponding theoretical predictions given by
Eq. (\ref{ecCociente}), finding an excellent agreement with the MD
simulations results for the whole range of inelasticities. This is
remarkable as there is not any fitting parameter. It can be
appreciated that the agreement in the $\epsilon=0.2$ case is better
than for $\epsilon=0.5$. This was expected since the theory is
implemented by a power expansion around $\epsilon=0$. The quasielastic
theoretical predictions of Ref. \cite{mgb19} for $\epsilon=0.2$ and
$\epsilon=0.5$ are also plotted (dashed 
lines). Although this prediction captures the tendency of
the data, it is clearly seen that the new prediction given by 
Eq. (\ref{ecCociente}) improves considerably the agreement with the
simulation results, specially for strong inelasticities. 

\begin{figure}
\begin{center}
\includegraphics[angle=0,width=0.7\linewidth,clip]{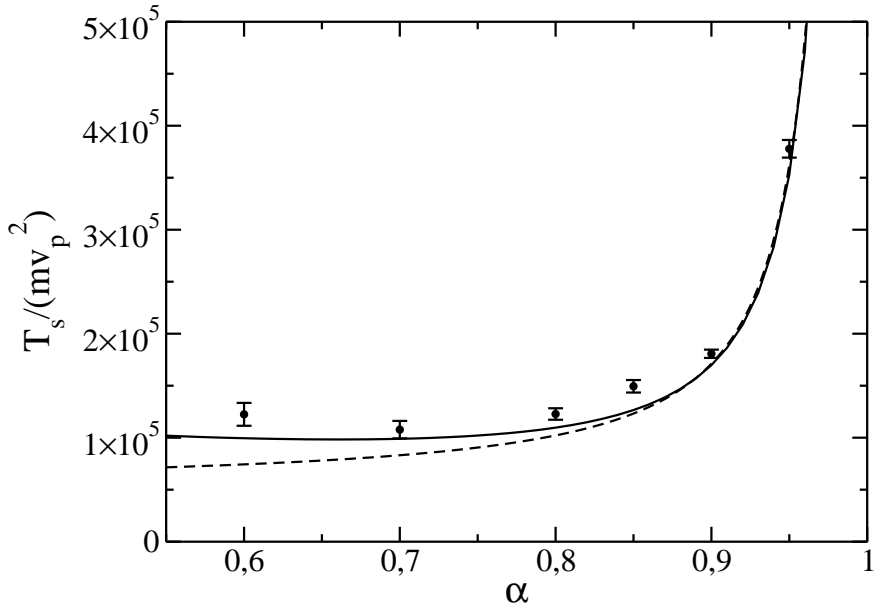}
\end{center}
\caption{Stationary horizontal temperature scaled with $mv_p^2$ for
  $\epsilon=0.5$. The circles are the simulation results and the
  solid line is the theoretical prediction given by
Eq. (\ref{ecTempSt}). The dashed line is the quasielastic
theoretical prediction of Ref. \cite{mgb19}.}\label{FigT05}
\end{figure}

\begin{figure}
\begin{center}
\includegraphics[angle=0,width=0.7\linewidth,clip]{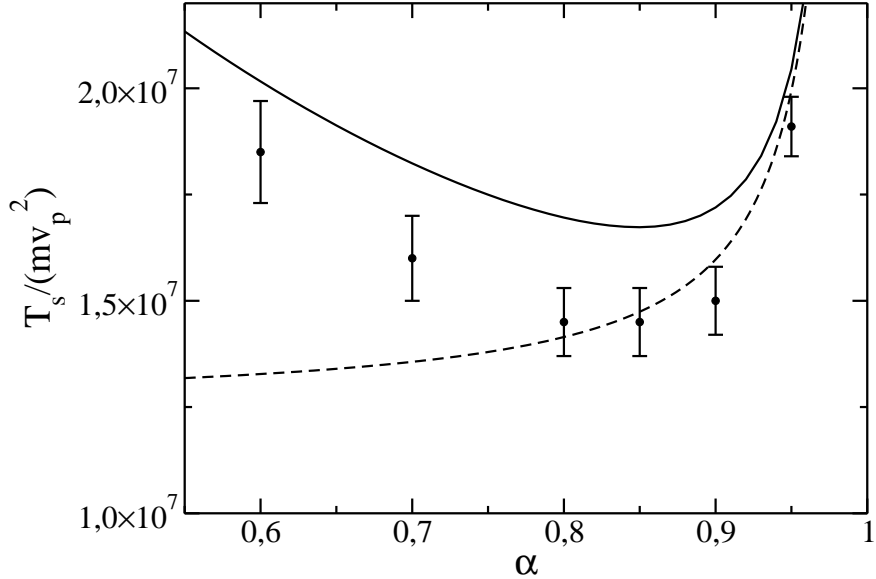}
\end{center}
\caption{Stationary horizontal temperature scaled with $mv_p^2$ for
  $\epsilon=0.2$. The circles are the simulation results and the
  solid line is the theoretical prediction given by
Eq. (\ref{ecTempSt}). The dashed line is the quasielastic
theoretical prediction of Ref. \cite{mgb19}.}\label{FigT02}
\end{figure}

In Fig. \ref{FigT05} we have plotted the stationary horizontal
temperature scaled with $mv_p^2$ for $\epsilon=0.5$ as a function of
$\alpha$. The circles are the MD simulation results and the solid line
the theoretical prediction given by Eq. (\ref{ecTempSt}). The dashed
line is the quasielastic prediction calculated in
Ref. \cite{mgb19}. The error bars are evaluated as in
Fig. \ref{FigCociente}. Again, the agreement between the theoretical
prediction and the simulation results is very good, 
and the ``inelastic'' prediction improves the agreement with respect
to the quasielastic one. Let us remark 
that the quasielastic prediction of the stationary horizontal
temperature decays monotonically with the inelasticity, while
Eq. (\ref{ecTempSt}) predicts an enhanced that actually is observed in the
simulations. The same is plotted in Fig. \ref{FigT02} for
$\epsilon=0.2$. Here, it seems that the agreement with the quasielastic
prediction is better than with Eq. (\ref{ecTempSt}) for
$0.8<\alpha<0.9$, although it is clear that this is not the case for
strong inelasticities. In fact, the minimum of the stationary
temperature measured in the simulation is around $\alpha\sim 0.85$ as
predicts Eq. (\ref{ecTempSt}). Let us also stress that
$\frac{T_s}{mv_p^2}\sim 10^5$ for $\epsilon=0.5$ and
$\frac{T_s}{mv_p^2}\sim 10^7$ for $\epsilon=0.2$, i.e. the thermal
horizontal velocity is much larger than the velocity of the wall. This
was commented in the previous section and it is a consequence of the
fact that there are two dimensionless parameter, $\epsilon$ and
$\tilde{n}\sigma^2$, that contribute to the increase of $T_s$ (see
Eq. (\ref{ecTempSt})). 

\begin{figure}
\begin{center}
\includegraphics[angle=0,width=0.7\linewidth,clip]{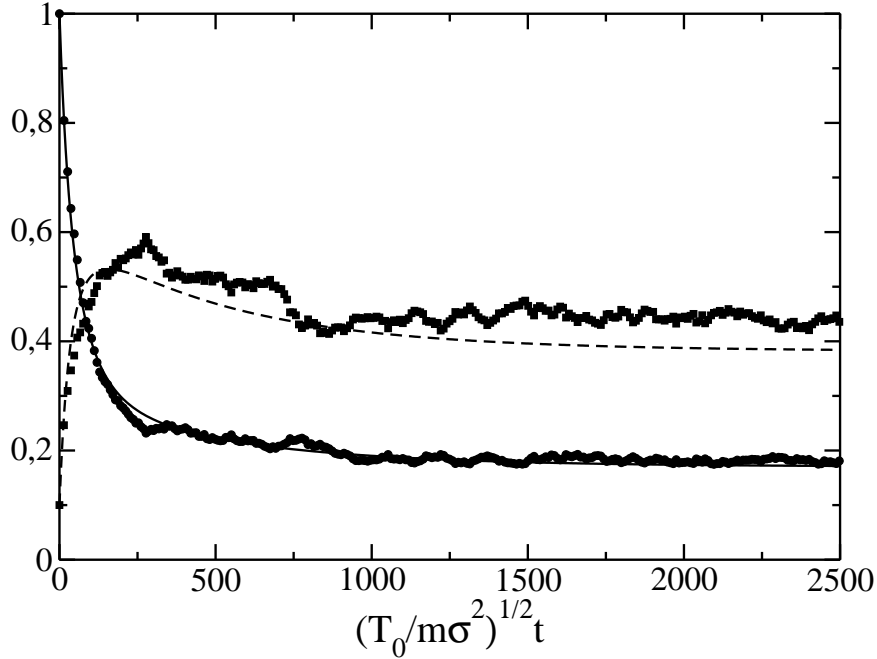}
\end{center}
\caption{ Time evolution of the
horizontal (circles) and vertical (squares) temperatures for
$\epsilon=0.5$ and $\alpha=0.9$. The solid and dashed lines are the numerical
solution of Eqs. (\ref{ecT}) and (\ref{ecTz}) for the horizontal and
vertical temperatures respectively.}\label{FigEdos}
\end{figure}

Let us examine now if Eqs. (\ref{ecT}) and (\ref{ecTz}) describe
correctly the time evolution of the system. In Fig. \ref{FigEdos} we
have plotted the MD simulation results of the time evolution of the
horizontal (circles) and vertical (squares) temperatures. The
dimensionless height of the system is $\epsilon=0.5$, $\alpha=0.9$, and
the initial condition was taken to be a Gaussian with
$T_z(0)=0.1T_0$. The solid and dashed lines are the numerical
solution of Eqs. (\ref{ecT}) and (\ref{ecTz}) for the horizontal and
vertical temperatures, respectively. The agreement with the simulation
results is excellent for the whole time evolution. Note that there is
a time window, around $(100,300)$ in the dimensionless time scale 
$\left[\frac{T_0}{m}\right]^{1/2}\frac{t}{\sigma}$, in which the
degree of freedom with the highest granular temperature (the vertical 
one) heats up while the degree of freedom with the lowest granular
temperature (the horizontal one)  cools down. A similar good agreement
between the MD simulation results and the theoretical predictions is 
obtained for other values of the parameters and initial conditions. 

\begin{figure}
\begin{center}
\includegraphics[angle=0,width=0.7\linewidth,clip]{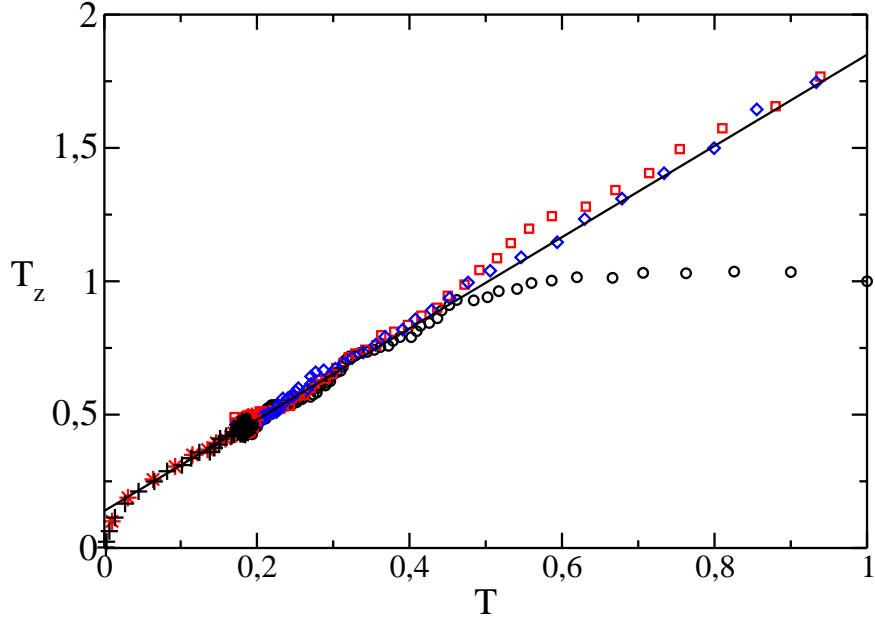}
\end{center}
\caption{ Vertical
temperature vs. horizontal temperature for
$\alpha=0.9$ and $\epsilon=0.5$. The initial conditions are (color
online) $(T(0),T_z(0))=(T_0, T_0)$ (black circles), $(T(0),
T_z(0))=(10T_0, 10T_0)$ (red squares), $(T(0),
T_z(0))=(20T_0,20T_0)$ (blue diamonds), $(T(0),
T_z(0))=(0.002T_0, 0.002T_0)$ (black pluses), and $(T(0),
T_z(0))=(0.01T_0, 0.01T_0)$ (red stars). $T_0$ is taken as unity. The solid line
is the theoretical prediction for the hydrodynamic regime close to the
stationary state. }\label{TzTFig} 
\end{figure}

For given values of the parameters, we have studied the dynamics of
the system starting with different initial conditions. The initial
condition was always a Gaussian with two temperatures. It is found
that, after a transient, the system reaches a regime in which the
horizontal and vertical temperatures are related, independently of the
initial condition. This is the generalization to the non-linear case
of the homogeneous hydrodynamic regime studied in the linear case in
the previous section. In effect, in Fig. \ref{TzTFig} the vertical
temperature is plotted as a function of the horizontal temperature for
$\alpha=0.9$ and $\epsilon=0.5$ 
starting with different initial conditions. The value of the velocity
of the wall is, in this case, the same for all the initial conditions,
$v_p=0.001\left(\frac{T_0}{m}\right)^{1/2}$, $T_0$ being the initial
horizontal temperature of one of the simulations that, as before, is
taken to be unity. The initial conditions are (color online) $(T(0),
T_z(0))=(T_0, T_0)$ (black circles), $(T(0),
T_z(0))=(10T_0, 10T_0)$ (red squares), $(T(0),
T_z(0))=(20T_0,20T_0)$ (blue diamonds), $(T(0),
T_z(0))=(0.002T_0, 0.002T_0)$ (black pluses), and $(T(0),
T_z(0))=(0.01T_0, 0.01T_0)$ (red stars). It is seen that, after a transient,
the curves collapse to a single curve through which the stationary
state is reached ($T_s\approx 0.18T_0$ and $T_{z,s}\approx
0.45T_0$). In the figure, the universal curve in the linear regime is
also plotted, $T_z^{(H)}(T)\approx T_{z,s}+q(T-T_s)$, with the values
of the stationary temperatures taken from the simulations and the
value of $q$ taken from
the theoretical prediction given by Eq. (\ref{ecP}). 
\begin{figure}
\begin{center}
\includegraphics[angle=0,width=0.7\linewidth,clip]{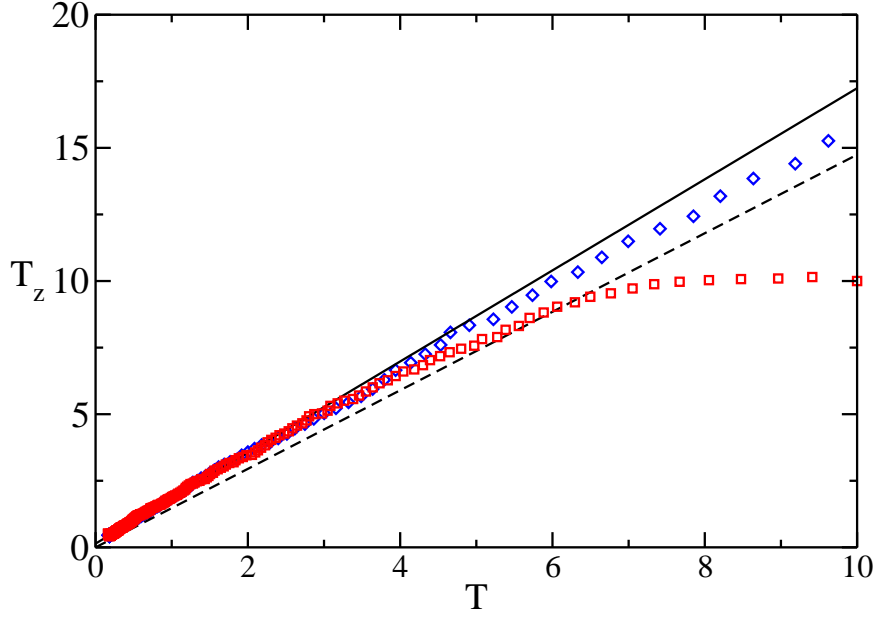}
\end{center}
\caption{ Vertical
temperature vs. horizontal temperature for
$\alpha=0.9$ and $\epsilon=0.5$. The initial conditions are
$(T(0),T_z(0)) =(10T_0, 10T_0)$ (red squares) and $(T(0),T_z(0))
=(20T_0, 20T_0)$ (blue diamonds). The solid and dashed lines are the
theoretical predictions close to the stationary state and for the free cooling
case respectively, both in the
hydrodynamic regime. }\label{TzT2Fig}
\end{figure}
As it can be seen, this curve is a good approximation for the region
in which the temperatures collapse (the universal homogeneous
hydrodynamic regime) even for temperatures that are far apart the
stationary state, and  for which the linear equations,
Eqs. (\ref{linearizedEqs}), are not supposed to be valid. 
Why $T_z^{(H)}(T)$ can be approximated by its
linear expansion around the stationary state out of the linear regime?
This can be understood, at least for $T>T_s$. In effect, for high
temperatures (compared to $T_s$), the system does not ``feel'' the
vibrating wall and may evolve cooling freely. In this case, the
dynamics of the temperatures is linear in the $s$ variable 
\begin{equation}\label{linearizedEqsFree}
\frac{d}{ds}\left[\begin{array}{c}
T\\
T_z
\end{array}\right]=M_f \left[\begin{array}{c}
T\\
T_z
\end{array}\right], 
\end{equation}
where we have introduced the matrix
\begin{equation}
M_f=\left[\begin{array}{cc}
-1+\alpha-\frac{5\alpha-1}{12}\epsilon^2 & \frac{3\alpha+1}{12}\epsilon^2\\
\frac{1+\alpha}{3}\epsilon^2 &
                                                             -\frac{2}{3}\epsilon^2 
\end{array}\right]. 
\end{equation}
The structure of the matrix $M_f$ is similar to $M$, in the sense that
the two eigenvalues are negative one of them dominating the dynamics
in the long time limit, so that $T_z\approx q_fT_z$ in this
regime \cite{bgm19}. The coefficient $q_f$ can be calculated in a similar fashion
that $q$ obtaining, in addition, that they are very close. Hence, coming back
to the vibrating case again, the universal curve goes from the free
cooling behavior for high temperatures to the linear regime close to
the stationary state. As the two functions are similar and the
transition from one case to the other is expected to be smooth, the
function $T_z^{(H)}(T)\approx T_{z,s}+q(T-T_s)$ is expected to be a
good approximation for $T_z^{(H)}(T)$ in a much wider regime. In
Fig. \ref{TzT2Fig} the results for the initial conditions
$(T(0),T_z(0)) =(10T_0, 10T_0)$ (red squares) and $(T(0),T_z(0)) 
=(20T_0, 20T_0)$ (blue diamonds) are plotted in a wider scaled, finding
that $T_z^{(H)}(T)\approx T_{z,s}+q(T-T_s)$ (solid line) is reliable
even till $T\sim 6T_0$. The universal curve in the free cooling case
is also plotted (dashed line) finding that, in effect, $q\sim
q_f$. Similar results are found for other values of the parameters. 

\begin{figure}
\begin{center}
\includegraphics[angle=0,width=0.7\linewidth,clip]{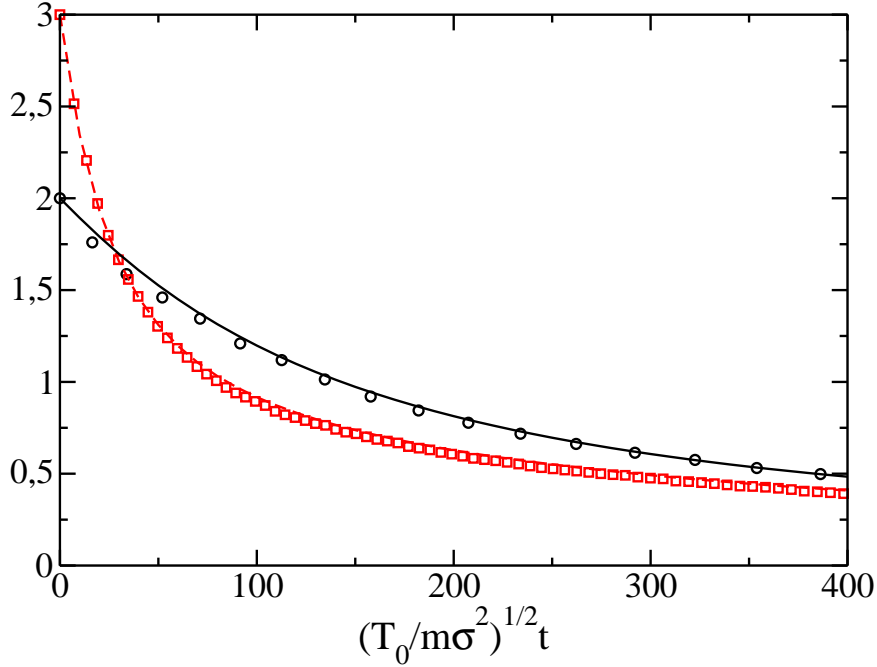}
\end{center}
\caption{ Time evolution of the horizontal temperature for
  $\epsilon=0.5$, $\alpha=0.9$ and two different 
initial conditions (color online), $(T_A(0),T_{z,A}(0))=(2T_0, 3.5T_0)$ (black
circles) and $(T_B(0),T_{z,B}(0))=(3T_0, 0.5T_0)$ (red squares). The
(black) solid line and (red) dashed line are their corresponding
theoretical predictions.}\label{mpembaFig}
\end{figure}

Finally, let us consider two systems, $A$ and $B$ with the same values
of the parameters, but prepared in such a way that $\delta T_B(0)>\delta
T_A(0)>0$ ($\delta T_B(0)<\delta
T_A(0)<0$) . Is it possible that, still, system $B$ reaches the
stationary state faster than
system $A$? In the linear regime, the question can be tackled with
Eq. (\ref{linearizedEqs}) by choosing the appropriate $\delta
T_{z,A}(0)$ and $\delta T_{z,B}(0)$. A necessary condition for the
effect is that the two curves cross each other at some time,
$s>0$, that occurs if 
\begin{equation}\label{mpeq}
e^{(\lambda_1-\lambda_2)s}=-\frac{u_{21}}{u_{11}}\frac{v_{21}+v_{22}\frac{\Delta T_z}{\Delta
    T}}
{v_{11}+v_{12}\frac{\Delta T_z}{\Delta T}}, 
\end{equation}
where we have introduced  $\Delta T\equiv\delta T_B(0)-\delta T_A(0)$
and $\Delta T_z\equiv\delta T_{z,B}(0)-\delta T_{z,A}(0)$. Note that
Eq. (\ref{mpeq}) depends only on one single parameter 
$\frac{\Delta T_z}{\Delta T}$. For the studied values of the
parameters, it is found that Eq. (\ref{mpeq}) has no solution for
$\frac{\Delta T_z}{\Delta T}>0$ and has one solution for $\frac{\Delta
  T_z}{\Delta T}<0$. In any case, even if the curves cross each other,
it can happen that the initially hotter system crosses the stationary
value and reaches the stationary state less quickly. By analyzing the
Eqs. (\ref{ecT}) and (\ref{ecTz}), it can be shown qualitatively that
the condition $\frac{\Delta T_z}{\Delta T}<0$ can work also in the
non-linear regime. This is because the above mentioned condition
implies that the hottest initial configuration 
will cool down much quicker than the coolest one. In
Fig. \ref{mpembaFig} the time evolution of the horizontal temperature
is plotted for $\epsilon=0.5$, $\alpha=0.9$ and two different
initial conditions (color online), $(T_A(0),T_{z,A}(0))=(2T_0, 3.5T_0)$ (black
circles) and $(T_B(0),T_{z,B}(0))=(3T_0, 0.5T_0)$ (red squares). The
(black) solid line and (red) dashed line are their corresponding
theoretical predictions, i.e. the numerical solution of
Eqs. (\ref{ecT}) and (\ref{ecTz}) with the
corresponding initial condition. It is $\frac{\Delta T_z}{\Delta
  T}<0$ and, in fact, the two curves cross each other. Note that we are
far from the linear regime as $T_s\approx 0.18T_0$ and $T_{z,s}\approx
0.45T_0$. A similar effect has been studied previously in the context
of granular systems \cite{lvps17, tllvps19}. In any case, it is
important to remark that the effect does not contradict the
hydrodynamic behavior as it occurs in the kinetic time scale. 

\section{Discussion and conclusions}\label{secConclusions}

In this paper we have proposed a closed dynamical equation for the
one-particle distribution function for a system of inelastic hard
spheres confined between two parallel hard flat plates valid in the
low density limit. The distance
between the walls is smaller than twice the diameter of the particles
so that the system is actually Q2D, and the bottom plate is a sawtooth
wall that injects energy into the system. The structure of the
equation is similar to the ``traditional'' Boltzmann equation: it
contains a free streaming contribution, a collisional term and a wall
term. The collisional contribution takes into account the effect of
the confinement because, given a tagged particle, only collisions in
some directions are possible. The wall contribution depends on the
nature of the wall, concretely on the wall-particle collision
rule. From the kinetic equation and assuming that the system is
spatially homogeneous, we have derived the evolution equations for the
horizontal and vertical temperatures. For the derivation, it has been
assumed that the one-particle distribution function is a Gaussian with
two temperatures (the horizontal and vertical ones) and that
$\epsilon\equiv\frac{H-\sigma}{\sigma}\ll 1$. A very good agreement 
between the theoretical predictions for the temperatures and MD
simulation results is found for a wide range of inelasticities if
$\epsilon\le 0.5$ without any fitting parameter. This good agreement
is obtained not only for the stationary 
values, but also for the whole dynamics of the temperatures. It has
been shown that, under certain conditions, the system with a larger
initial granular temperature cools down more quickly
and reaches the stationary state before than another system with an
initial condition closer to the stationary state. Actually, it seems
that the experimental realization of this effect is easy in a vibrated
granular monolayer. The reason is that the relevant parameters are the 
horizontal and vertical temperatures that are easily controlled
experimentally (the
vertical temperature can be changed just by changing the
parameters of the vibrating wall). Other memory effects such as the
Kovacs effect \cite{k63} can also be studied in the context of our model. In
contrast to other granular systems \cite{pt14}, the horizontal
temperature always has an anomalous behavior because the horizontal
temperature equation does not contain $v_p$ and a sudden change of it
does not change the cooling/heating rate.  

Remarkably, it is found that, independently of the
initial condition, after a transient, the system reaches a universal
regime in which the two temperatures are related, i.e. the vertical
temperature is a function of the horizontal temperature, $T_z^{(H)}(T)$, or
vice versa, $T^{(H)}(T_z)$. This is the so-called homogeneous
hydrodynamic regime. Numerical simulation results show that, in this
regime, the relation between the temperatures
is approximately linear and can be written in the form,
$T_z^{(H)}(T)\approx T_{z,s}+p(T-T_s)$, even for 
temperatures for which the linearized equations are not expect to be
reliable. Hence, in the homogeneous hydrodynamic regime, the following
approximated closed equation for the horizontal temperature is
obtained
\begin{eqnarray}
\frac{dT}{ds}&=&\left[-(1-\alpha)
+\frac{\epsilon^2}{12}\left[-(5\alpha-1)+(3\alpha+1)p\right]\right]T
+\frac{\epsilon^2(3\alpha+1)}{12}T_s. 
\end{eqnarray}
The study of the homogeneous hydrodynamic regime is relevant, as it is
the first step to be done for the ulterior study of hydrodynamic in
the plane. Here, by hydrodynamic in the plane we mean a closed
description of the system in terms of the local two-dimensional density,
projected flow velocity on the plane and horizontal
temperature. Actually, in the present model, the study of homogeneous
hydrodynamic is specially relevant from a quantitative point of view
as compared to other models. In effect, in the stochastic thermostat
model or in the $\Delta$ model the one-particle distribution function
in the homogeneous hydrodynamic regime is always close to a Gaussian,
the deviation from it described by the kurtosis, $a_2$, that is always
very small \cite{gmt12, bmgb14}. The transport
coefficients, as calculated by the Chapman-Enskog scheme or by
linear response methods, 
depend on the dynamics of the 
one-particle distribution function in the homogeneous hydrodynamic
regime and, specifically, on the dynamics of $a_2$ \cite{gmt13, gcv13, bbmg15}. As
$a_2\ll 1$, its effect on the transport 
coefficients is also very small and the Gaussian approximation (with
one temperature) is a good approximation. In contrast, in our model there is not
a small parameter in the one-particle distribution function, the
one-temperature Gaussian approximation is not a good approximation and
the effects of homogeneous hydrodynamics encoded in $T_z^{(H)}(T)$ are
expected to be relevant for the computation of the transport
coefficients. 

Finally, let us mention that the kinetic equation can be extended to
higher densities by using the Enskog approximation, i.e. assuming that
there are not velocity correlation between colliding particles,
although spatial correlations are taken into account through the pair
correlation function at contact \cite{mgb18}. The situation in this
case is more complex as it is not 
clear that the system can be considered to be homogeneous in the
vertical direction and, in addition, it is possible that the
dependence on the orientation of the pair correlation function be
relevant for the analysis. Work in these lines is in progress.

\section{Acknowledgments}

This research was supported by the Ministerio de Educaci\'{o}n,
Industria y Competitividad (Spain) through Grant No. FIS2017-87117-P (partially financed
by FEDER funds). 

\appendix

\section{Evaluation of some collisional integrals}\label{cf}
The objective of the Appendix is to evaluate the collisional integrals
that appears in the equations of the temperatures in the
two-temperatures Gaussian approximation given by
Eq. (\ref{fapprox}). The following property of the collision operator
will be used
\begin{eqnarray}\label{propJz}
&&\int_{\sigma/2}^{H-\sigma/2} dz\int d\mathbf{v}\psi(\mathbf{v})J_z[f|f] \nonumber\\
&&=\frac{\sigma^2}{2}\int
d\mathbf{v}_1\int
d\mathbf{v}_2f(\mathbf{v}_1)f(\mathbf{v}_2)\int_{\sigma/2}^{H-\sigma/2}
dz\int_{\Omega(z)}
d\sig\Theta(\mathbf{v}_{12}\cdot\sig)|\mathbf{v}_{12}\cdot\sig|(b_{\sig}-1)
[\psi(\mathbf{v}_1)+\psi(\mathbf{v}_2)].  \nonumber\\
\end{eqnarray}
Taking into account Eq. (\ref{propJz}), the integral that appears in
the vertical temperature equation is 
\begin{eqnarray}\label{vxvyjz}
&&\frac{1}{H-\sigma}\int_{\sigma/2}^{H-\sigma/2} dz\int
   d\mathbf{v}(v_x^2+v_y^2)J_z[f|f]
\nonumber\\
&&=\frac{\sigma^2}{2(H-\sigma)}\int
d\mathbf{v}_1\int
d\mathbf{v}_2f(\mathbf{v}_1)f(\mathbf{v}_2)\int_{\sigma/2}^{H-\sigma/2}
dz\int_{\Omega(z)}
d\sig\Theta(\mathbf{v}_{12}\cdot\sig)|\mathbf{v}_{12}\cdot\sig|(b_{\sig}-1)
(v_{1x}^2+v_{1y}^2+v_{2x}^2+v_{2y}^2). \nonumber\\
\end{eqnarray}
And using the collision rule, Eqs (\ref{cr1}) and (\ref{cr2}), we have
\begin{eqnarray}
(b_{\sig}-1)
(v_{1x}^2+v_{1y}^2+v_{2x}^2+v_{2y}^2)=-(1+\alpha)(\sig\cdot\mathbf{g})(g_x\hat{\sigma}_x
+g_y\hat{\sigma_y})+\frac{(1+\alpha)^2}{2}(\sig\cdot\mathbf{g})^2
(\hat{\sigma}_x^2+\hat{\sigma}_y^2), \nonumber\\
\end{eqnarray}
where we changed the notation of the relative velocity,
$\mathbf{g}\equiv\mathbf{v}_{12}$, for simplicity. Then,
Eq. (\ref{vxvyjz}) can be written as 
\begin{eqnarray}
\frac{1}{H-\sigma}\int_{\sigma/2}^{H-\sigma/2} dz\int
d\mathbf{v}(v_x^2+v_y^2)J_z[f|f]=
\frac{\sigma^2}{2(H-\sigma)}\int d\mathbf{v}_1\int
  d\mathbf{v}_2f(\mathbf{v}_1)f(\mathbf{v}_2)\int_{\sigma/2}^{H-\sigma/2}dz\nonumber\\
\int_{\Omega(z)}d\sig
\Theta(\mathbf{g}\cdot\sig)|\mathbf{g}\cdot\sig|
\left[-(1+\alpha)(\sig\cdot\mathbf{g})(g_x\hat{\sigma}_x
+g_y\hat{\sigma_y})+\frac{(1+\alpha)^2}{2}(\sig\cdot\mathbf{g})^2
(\hat{\sigma}_x^2+\hat{\sigma}_y^2)\right]\nonumber\\
=\frac{\sigma^2}{2(H-\sigma)}\int d\mathbf{v}_1\int
  d\mathbf{v}_2f(\mathbf{v}_1)f(\mathbf{v}_2)\int_{\sigma/2}^{H-\sigma/2}dz\nonumber\\
\int_{\Omega(z)}d\sig
\Theta(-\mathbf{g}\cdot\sig)|\mathbf{g}\cdot\sig|
\left[-(1+\alpha)(\sig\cdot\mathbf{g})(g_x\hat{\sigma}_x
+g_y\hat{\sigma_y})+\frac{(1+\alpha)^2}{2}(\sig\cdot\mathbf{g})^2
(\hat{\sigma}_x^2+\hat{\sigma}_y^2)\right], \nonumber\\
\end{eqnarray}
where we have changed $\mathbf{v}_1$ by $\mathbf{v}_2$ in the last
step. Hence, we can get rid of the $\Theta$ function and we have
\begin{eqnarray}\label{h1h2}
\frac{1}{H-\sigma}\int_{\sigma/2}^{H-\sigma/2} dz\int
d\mathbf{v}(v_x^2+v_y^2)J_z[f|f]=\frac{1}{H-\sigma}\int_{\sigma/2}^{H-\sigma/2}dz
\left[\frac{(1+\alpha)^2}{2}H_1(z)-(1+\alpha)H_2(z)\right], \nonumber\\
\end{eqnarray}
where 
\begin{eqnarray}
H_1(z)&=&\frac{\sigma^2}{4}\int d\mathbf{v}_1\int
  d\mathbf{v}_2f(\mathbf{v}_1)f(\mathbf{v}_2)
\int_{\Omega(z)}d\sig|\mathbf{g}\cdot\sig|^3
(\hat{\sigma}_x^2+\hat{\sigma}_y^2), \\
H_2(z)&=&\frac{\sigma^2}{4}\int d\mathbf{v}_1\int
  d\mathbf{v}_2f(\mathbf{v}_1)f(\mathbf{v}_2)
\int_{\Omega(z)}d\sig|\mathbf{g}\cdot\sig|(\mathbf{g}\cdot\sig)
(g_x\hat{\sigma}_x+g_y\hat{\sigma}_y). 
\end{eqnarray}

Let us first calculate $H_1$. To do it, let us perform the following
change of variables
\begin{eqnarray}\label{varC}
\mathbf{C}&=&\frac{1}{2}(\mathbf{c}_1+\mathbf{c}_2), \\
\mathbf{c}&=&\mathbf{c}_1-\mathbf{c}_2, \label{varc}
\end{eqnarray}
where
\begin{equation}
\mathbf{c}_i=\frac{1}{w}(v_{i,x}\mathbf{e}_x+v_{i,y}\mathbf{e}_y)+\frac{v_{i,z}}{w_z}\mathbf{e}_z, 
\end{equation}
for $i=1,2$.  Taking into account the expression of the distribution function given
by Eq. (\ref{fapprox}) 
\begin{equation}
f(\mathbf{v}_1)f(\mathbf{v}_2)=\frac{n^2}{\pi^3w^4w_z^2}e^{-2C^2}e^{-\frac{c^2}{2}}, 
\end{equation}
we have
\begin{eqnarray}
H_1(z)=\frac{n^2\sigma^2}{4\pi^3}\int d\mathbf{C}e^{-2C^2}\int
  d\mathbf{c}e^{-\frac{c^2}{2}}
\int_{\Omega(z)}d\sig|\mathbf{c}\cdot\mathbf{a}|^3(w^2\hat{\sigma}_x^2+
  w^2\hat{\sigma}_y^2+w_z^2\hat{\sigma}_z^2)^{3/2}(\hat{\sigma}_x^2+\hat{\sigma}_y^2), 
\nonumber\\
\end{eqnarray}
where we have introduced the unit vector 
\begin{equation}\label{unitVectora}
\mathbf{a}\equiv\frac{w\hat{\sigma}_x\mathbf{e}_x+
  w\hat{\sigma}_y\mathbf{e}_y
+w_z\hat{\sigma}_z\mathbf{e}_z}{\sqrt{w^2\hat{\sigma}_x^2+
  w^2\hat{\sigma}_y^2+w_z^2\hat{\sigma}_z^2}}. 
\end{equation}
The velocity integrals in $\mathbf{C}$ and $\mathbf{c}$ are
\begin{eqnarray}
\int d\mathbf{C}e^{-2C^2}=\frac{\pi^{3/2}}{2\sqrt{2}}, \quad
\int d\mathbf{c}e^{-\frac{c^2}{2}}|c_x|^3=8\pi, 
\end{eqnarray}
so that, by symmetry, it is
\begin{equation}
H_1(z)=\frac{n^2\sigma^2}{\sqrt{2\pi}}\int_{\Omega(z)}d\sig(w^2\hat{\sigma}_x^2+
  w^2\hat{\sigma}_y^2+w_z^2\hat{\sigma}_z^2)^{3/2}(\hat{\sigma}_x^2+\hat{\sigma}_y^2), 
\end{equation}
or, by introducing the parametrization of $\Omega(z)$ given by
Eq. (\ref{omegaPar})
\begin{equation}
H_1(z)=\sqrt{2\pi}
n^2\sigma^2\int_{\frac{\pi}{2}-b_2(z)}^{\frac{\pi}{2}+b_1(z)}d\theta\sin^3\theta
(w^2\sin^2\theta+w_z^2\cos^2\theta)^{3/2}.  
\end{equation}
Changing variables to $z_1$ defined by Eq. (\ref{thetaz1}) and integrating also in
$z$, we obtain
\begin{eqnarray}
\int_{\sigma/2}^{H-\sigma/2}dzH_1(z)
=\sqrt{2\pi}n^2\sigma\int_{\sigma/2}^{H-\sigma/2}dz
\int_{\sigma/2}^{H-\sigma/2}dz_1\left[1-\left(\frac{z_1-z}{\sigma}\right)^2\right]\nonumber\\
\left\{w^2\left[1-\left(\frac{z_1-z}{\sigma}\right)^2\right]
+w_z^2 \left(\frac{z_1-z}{\sigma}\right)^2\right\}^{3/2}, 
\end{eqnarray}
or, in terms of the dimensionless variables
\begin{equation}
\tilde{z}_1=\frac{z_1-\sigma/2}{\sigma}, \quad \tilde{z}_2=\frac{z-\sigma/2}{\sigma}, 
\end{equation}
\begin{eqnarray}
\int_{\sigma/2}^{H-\sigma/2}dzH_1(z)
=2\sqrt{2\pi}n^2\sigma^3\int_0^\epsilon d\tilde{z}_1
\int_0^{\tilde{z}_1}d\tilde{z}_2(1-\tilde{z}_{12})^2[w^2(1-\tilde{z}_{12}^2)
+w_z^2\tilde{z}_{12}^2]^{3/2}, 
\end{eqnarray}
where we have used that the integrand is invariant under the change
$\tilde{z}_1$ by $\tilde{z}_2$ and we have introduced 
$\tilde{z}_{12}\equiv \tilde{z_1}-\tilde{z}_2$. Finally, changing variables to 
\begin{equation}
y=\tilde{z_1}-\tilde{z}_2, \quad
Y=\frac{1}{2}(\tilde{z_1}+\tilde{z}_2), 
\end{equation}
it is obtained
\begin{eqnarray}
\int_{\sigma/2}^{H-\sigma/2}dzH_1(z)
=2\sqrt{2\pi}n^2\sigma^3\int_0^\epsilon dy
\int_{y/2}^{\epsilon-y/2}dY
  (1-y^2)[w^2(1-y^2)+w_z^2y^2]^{3/2}\nonumber\\ \label{ecH1}
= 2\sqrt{2\pi}n^2\sigma^3\int_0^\epsilon dy(\epsilon-y)
  (1-y^2)[w^2(1-y^2)+w_z^2y^2]^{3/2}. 
\end{eqnarray}

The evaluation of $H_2(z)$ follows similar lines. Changing 
variables to $\mathbf{C}$ and $\mathbf{c}$ introduced in Eqs. (\ref{varC})
and (\ref{varc}) and integrating in $\mathbf{C}$, we have
\begin{eqnarray}
H_2(z)=\frac{n^2\sigma^2w}{8\sqrt{2}\pi^{3/2}}\int
  d\mathbf{c}e^{-\frac{c^2}{2}}
\int_{\Omega(z)}d\sig|\mathbf{c}\cdot\mathbf{a}|(\mathbf{c}\cdot\mathbf{a})
(w^2\hat{\sigma}_x^2+  w^2\hat{\sigma}_y^2+w_z^2\hat{\sigma}_z^2)
(c_x\hat{\sigma}_x+c_y\hat{\sigma}_y).  
\nonumber\\
\end{eqnarray}
Expressing $c_x\hat{\sigma}_x+c_y\hat{\sigma}_y$ in terms of the components of
$\mathbf{c}$ in the orthonormal basis 
$\{\mathbf{a}, \mathbf{b}_1,\mathbf{b}_2\}$ and integrating in $\mathbf{c}$, we have
\begin{equation}
H_2(z)=\frac{n^2\sigma^2w}{\sqrt{2\pi}}\int_{\Omega(z)}d\sig 
(w^2\hat{\sigma}_x^2+w^2\hat{\sigma}_y^2+w_z^2\hat{\sigma}_z^2)
(\hat{\sigma}_x\mathbf{e}_x\cdot\mathbf{a}+\hat{\sigma}_y\mathbf{e}_y\cdot\mathbf{a}), 
\end{equation}
where we have used that, by symmetry, $\int 
d\mathbf{c}e^{-\frac{c^2}{2}}
|\mathbf{c}\cdot\mathbf{a}|(\mathbf{c}\cdot\mathbf{a})(\mathbf{c}\cdot\mathbf{b}_i)=0$,
for $i=1,2$. Now, writing the explicit expression of
$\mathbf{e}_i\cdot\mathbf{a}$ with the aid of Eq.
(\ref{unitVectora}), we have
\begin{equation}
H_2(z)=\frac{n^2\sigma^2w^2}{\sqrt{2\pi}}\int_{\Omega(z)}d\sig
(w^2\hat{\sigma}_x^2+w^2\hat{\sigma}_y^2+w_z^2\hat{\sigma}_z^2)
(\hat{\sigma}_x^2+\hat{\sigma}_y^2). 
\end{equation}
Finally, following exactly the same steps that above, it is obtained
\begin{eqnarray}\label{ecH2}
\int_{\sigma/2}^{H-\sigma/2}dzH_2(z)
=2\sqrt{2\pi}n^2\sigma^3w^2\int_0^\epsilon dy
(\epsilon-y)(1-y^2)\sqrt{w^2(1-y^2)+w_z^2y^2}. 
\end{eqnarray}

Using Eqs. (\ref{h1h2}), (\ref{ecH1}) and (\ref{ecH2}), the expression for 
$\frac{1}{H-\sigma}\int_{\sigma/2}^{H-\sigma/2}dz\int
d\mathbf{v}\frac{m}{2}(v_x^2+v_y^2)J_z[f|f]$ given
by Eq. (\ref{ecTemxy}) is obtained. The evaluation of 
$\frac{1}{H-\sigma}\int_{\sigma/2}^{H-\sigma/2}dz\int
d\mathbf{v}\frac{m}{2}v_z^2J_z[f|f]$ follows exactly the same steps. 

Finally, the particle-wall collisional integral is 
\begin{eqnarray}
\frac{m}{2(H-\sigma)}\int
d\mathbf{v}v_z^2L_sf(\mathbf{v})=\frac{m}{2(H-\sigma)}\int
  d\mathbf{v}\theta(-v_z)|v_z|(4v_p^2-4v_pv_z)f(\mathbf{v})\nonumber\\
=\frac{m}{2(H-\sigma)}nv_pw_z\left(\frac{2}{\sqrt{\pi}}v_p+w_z\right), 
\end{eqnarray}
where the explicit form of the velocity distribution given by Eq. (\ref{fapprox})
has been used. Taking into account that, in order to be valid the
homogeneous approximation, the condition $v_p\ll w_z$ has to be
fulfilled, we obtain the expression of the main text.

\end{document}